\documentclass[twocolumn]{aastex63}
\usepackage{xcolor}
\usepackage{makecell}
\usepackage{pbox}
\usepackage{lineno}

\usepackage{amsmath}

\begin{document}

\title{Stable mass transfer can explain massive binary black hole mergers with a high spin component}

\author{Yong Shao}
\affiliation{Department of Astronomy, Nanjing University, Nanjing 210023, People's Republic of China; shaoyong@nju.edu.cn}
\affiliation{Key laboratory of Modern Astronomy and Astrophysics (Nanjing University), Ministry of
Education, Nanjing 210023, People's Republic of China; lixd@nju.edu.cn}

\author{ Xiang-Dong Li}
\affiliation{Department of Astronomy, Nanjing University, Nanjing 210023, People's Republic of China; shaoyong@nju.edu.cn}
\affiliation{Key laboratory of Modern Astronomy and Astrophysics (Nanjing University), Ministry of
Education, Nanjing 210023, People's Republic of China; lixd@nju.edu.cn}

\begin{abstract}

Recent gravitational wave observations showed that binary black hole (BBH) mergers with massive components are more likely to 
have high effective spins. In the model of isolated binary evolution, BH spins mainly originate from the angular 
momenta of the collapsing cores before BH formation. Both observations and theories indicate that BHs tend to possess relatively
low spins, the origin of fast-spinning BHs remains a puzzle. We investigate an alternative process
that stable Case A mass transfer may significantly increase BH spins during the evolution of massive BH binaries. 
We present detailed binary evolution calculations and find that this process can explain observed high spins of some
massive BBH mergers under the assumption of mildly super-Eddington accretion.

\end{abstract}


\section{Introduction}

Since the discovery of the first gravitational wave source GW150914 \citep{ab16}, there are about 90 
binary black hole (BBH) mergers reported to date \citep{ab19,aaa21a,aaa21b,nkw21}. 
A number of formation channels have been put forward to explain the origin of BBH mergers \citep[see][for a review]{mb22}.
In the isolated binary evolution channel, compact BH binaries are formed either through common envelope
evolution \citep[e.g.,][]{ty93,lp97,vt03,bh16,es16,sv17,km18,ktl18,mg18,gm18,sm19,bc20,zs20,bbs21} or through
stable mass transfer between the BH and its companion \citep[e.g.,][]{vdh17,nvs19,bf21,obi21,sl21,ggm21}. Alternatively, 
merging BHs can be formed via dynamical 
interactions in globular clusters \citep{db10,rh16,as17,pw19,ky20} or young stellar clusters 
\citep{zm14,dgm19,sm20,rm20,mbs22}. Other formation channels involve isolated multiple systems 
\citep{st17,ll18,hn18,fl19}, the chemically homogeneous 
evolution for rapidly rotating stars \citep{md16,dm16,ml16},  the disk of active galactic nuclei \citep{ar16,sm17,mf18}, 
as well as the evolution of Population III binary stars \citep{ki14,ts21}. 

Gravitational wave observations \citep{aaa21b} indicate that 
the majority of BBH mergers have low effective spin parameters
\begin{eqnarray}
\chi_{\rm eff} = \frac{M_{1,\rm BH}a_{1,\rm BH}\cos \theta_{1}+M_{2,\rm BH}a_{2,\rm BH}\cos \theta_{2}}{M_{1,\rm BH}+M_{2,\rm BH}},
\end{eqnarray}
with $  \chi_{\rm eff} \sim 0$. Here $ M_{1,\rm BH} $ and $ M_{2,\rm BH} $ are the masses of both components, $ a_{1,\rm BH} $
and $ a_{2,\rm BH} $ the dimensionless BH spin magnitudes, and $ \theta_{1} $ and $ \theta_{2} $ the angles 
made by each component spin relative to the binary orbital angular momentum. For other minority mergers, there is a feature
that the systems with the most extreme spins have heavier masses \citep[][see also Figure 3]{aaa21b}. 
The distribution of BH spin orientations offers vital clues on the evolutionary pathways that produce 
merging BBHs \citep[e.g.,][]{rco21,gtn21,chn21,s22}. BBH mergers formed from isolated binary evolution tend to have spins 
preferentially aligned with their orbital angular momenta, while dynamically assembled systems are expected to possess 
isotropically oriented spins. Recent analyses on the distribution of BH spin orientations supported the hypothesis that
all merging systems originate from the isolated binary evolution channel \citep{gtn21}. 
In this channel, tidal interaction between both components of BBH's progenitor systems is the most promising 
mechanism for the origin of BH spins, which is effective for 
the binaries with very close orbits \citep[e.g.,][]{qfm18,bkf20,bfq20,bf21,ob21,fl22}. 
More recently, \citet{vdc21} suggested that the stable mass-transfer channel mainly produces BBH mergers with 
component masses above $ 30 M_{\odot} $, while the common-envelope channel predominately forms systems with 
component masses below about $ 30 M_{\odot} $. Thus, we propose that the observed high effective spins of massive 
BBH mergers may relate to the process of mass accretion onto BHs during previous stable mass-transfer phases.
It is worth to note that tidal interaction \citep[e.g.,][]{ob21} may spin up 
only the second-formed BH (with rare case of equal mass ratios when both BHs may be spun up). In case of our study,
the first-born BH is a subject of spin-up which is important in context of e.g. X-ray binary observations \citep{prh03}. 

In this paper, we investigate the influence of Case A mass transfer on the evolution of the accreting BHs and 
show that likely accounts for BBH systems with a fast-spinning component.
Previous population synthesis studies
did not predict the formation of high spin BHs via mass accretion because this process was thought to occur vary rapidly
\citep[e.g.,][]{bf21,zb22} and BHs 
can hardly accrete any material during the evolution \citep[see e.g.,][as an exception]{vdc20}. 
However, more detailed investigations indicate efficient spin up of BHs  
by stable mass accretion in X-ray binaries \citep{kk99,prh03,fm15,sl20}. 
Evolution of close
massive primordial binaries has been well studied over past decades \citep[e.g.,][]{p94,wlb01,sl16}. If the initial orbital period is around
a few days, Roche lobe overflow starts with Case A mass transfer when the donor star still undergoes core hydrogen burning.
The Case A mass transfer episode usually consists of two phases. The first is a rapid phase during which the mass ratio of
binary components is more or less reversed. The rapid mass-transfer phase happens roughly on the thermal timescale of 
the donor star. This phase is followed by a slower mass-transfer phase which occurs on the nuclear-expansion timescale of 
the donor star. Compared to the rapid phase, mass transfer usually takes place at a rate several orders of magnitude
smaller in the slow phase. Since the population synthesis method simulates mass transfer via Roche lobe overflow
in a rather crude way, here we perform detailed binary evolution calculations to examine the possible influence of 
mass accretion on the BH spins.

\begin{figure*}[hbtp]
\centering
\includegraphics[width=0.85\textwidth]{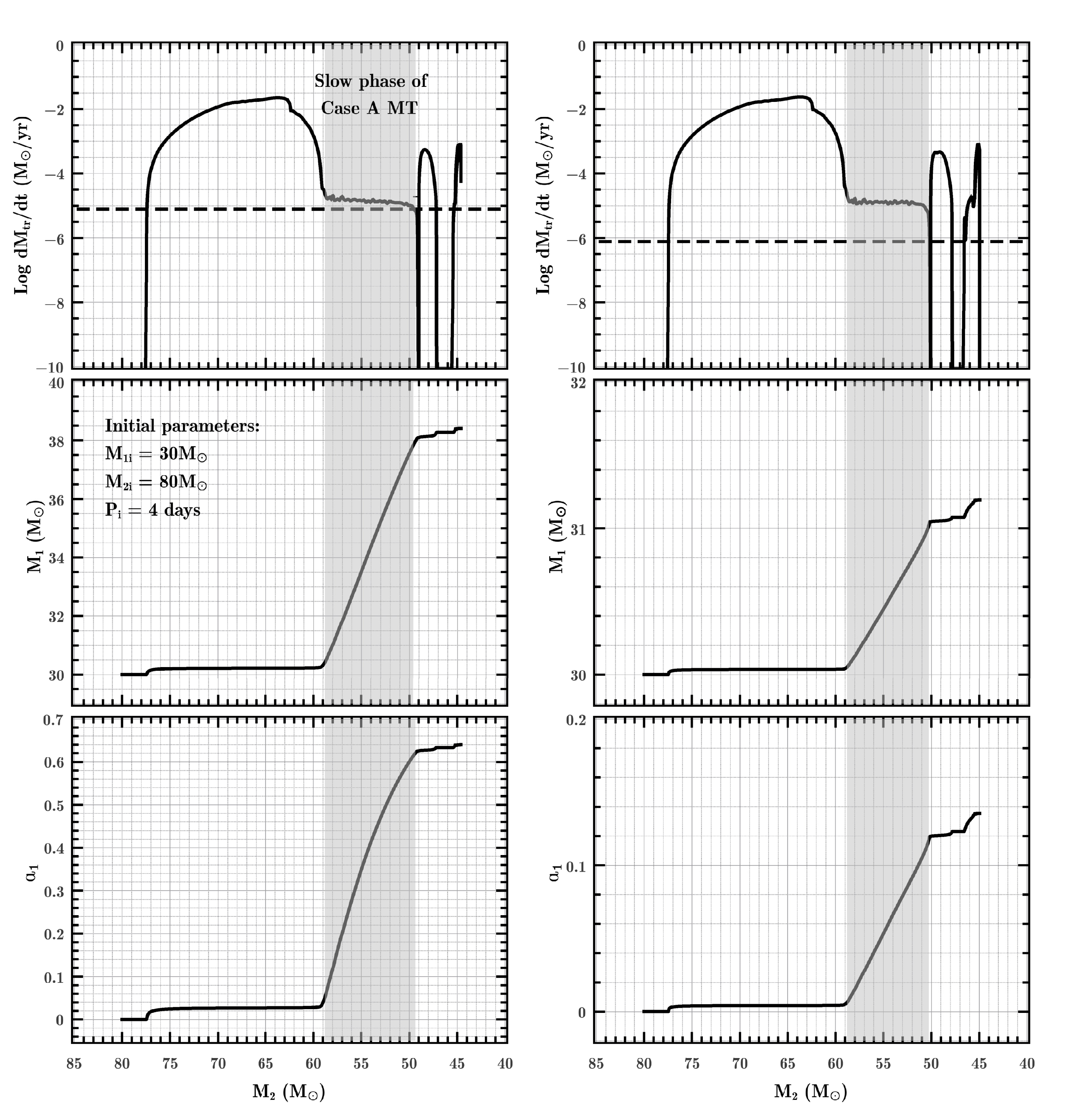}
\caption{Evolutionary tracks of a binary initially containing a $ 30M_{\odot} $ BH and a $ 80M_{\odot} $ donor in a 4-day orbit. 
The top, middle,
and bottom panels correspond to the mass transfer rate $ \dot{M}_{\rm tr} $, the BH mass $ M_{1} $, and its spin magnitude $ a_{1} $, 
respectively, as a function of the donor mass $ M_{2} $.  The left and right panels correspond to the BH accretion rate being limited by 
$\dot{\mathcal{M}}_{\rm Edd} $ and $  \dot{M}_{\rm Edd} $ during the evolution, respectively. The two dashed lines 
mark the positions of $\dot{\mathcal{M}}_{\rm Edd} $ and $  \dot{M}_{\rm Edd} $ for a $ 30M_{\odot} $ accreting BH.
The gray rectangle in each panel denotes the binary being in the slow phase of Case A mass transfer. 
   \label{figure1}}
\end{figure*}

\section{Method}

We use the  stellar evolution code Modules for Experiments in Stellar Astrophysics \textit{MESA} 
\citep[version 10398,][]{p11,p13,p15,pb18,pb19} to model the evolution of binary systems 
containing a donor star of mass $ 20-100 M_{\odot}$ (by a step of $ 20 M_{\odot}$) and an accreting BH of
mass $10-30M_{\odot} $ (by a step of $ 10 M_{\odot}$). The initial
binaries are set to have circular orbits, and the orbital periods increase from 1 day to 100 days by a 
logarithmic step of 0.1. The BH is regarded as a point mass, and the donor star begins its evolution from 
zero-age main sequence. All the models are calculated at solar metallicity ($ Z=0.02 $) and a
subsolar metallicity ($ Z=0.001 $). We refer to
\citet{sl21} for a detailed description of the input parameters in the code. 

For spherically symmetric accretion, the mass increase onto a BH of mass $ M_{1} $ is constrained by the Eddington limit 
\begin{eqnarray}
 \dot{M}_{\rm Edd} = \frac{4\pi G M_{1}}{\eta\kappa c} 
  \simeq  7.8\times 10^{-7} \left( \frac{M_{1}}{30M_{\odot}} \right)   M_{\odot}\rm yr^{-1}, 
\end{eqnarray}
where $ G $ is the gravitational constant, $ \kappa $ is the radiative opacity which taken to be $ 0.2(1+X) \,\rm cm^2\,g^{-1}$ 
for a composition with hydrogen mass fraction $ X $, 
$ c $ is the speed of light in vacuum, 
and $ \eta $ is the efficiency of the BH in converting rest mass into radiative energy \citep[see also][]{prh03}. This 
efficiency can be approximately given by 
\begin{eqnarray}
\eta = 1-\sqrt{1- \left( \frac{M_{1}}{3M_{\rm 1i} } \right)^{2}} 
\end{eqnarray}
for $ M_{1} < \sqrt{6} M_{\rm 1i}$, where  $ M_{\rm 1i} $ is the initial mass 
of the BH \citep{b70}.  Note that $ \eta \sim 0.1 $ always holds in our calculations. When the BH accretes 
mass and angular momentum, its spin magnitude $ a_{1} $ evolves according to 
\begin{eqnarray}
a_{1} = \left(  \frac{2}{3}\right)^{1/2} \frac{M_{\rm 1i}}{M_{\rm 1} } \left\lbrace 4-
\left[ 18\left(\frac{M_{\rm 1i}}{M_{1} }\right) ^{2}-2\right] ^{1/2} \right\rbrace 
\end{eqnarray}
for $ M_{1} < \sqrt{6} M_{\rm 1i}$ \citep{t74}.

For radiation pressure dominated accretion disks, \citet{b02} found that super-Eddington accretion rates of a factor
of 10 can be achieved due to the development of a photon-bubble instability \citep[see also][]{rb03}. This instability results in a large part 
of the disk volume being constituted of tenuous plasma, while the bulk of the mass is constrained in high-density regions.
The photons can diffuse out of the disk mostly through the tenuous regions, therefore enhancing the Eddington limit. 
In our calculations, we relax the Eddington accretion rate to be 
\begin{eqnarray}
 \dot{\mathcal{M}}_{\rm Edd} = 10 \dot{M}_{\rm Edd} 
  \simeq  7.8\times 10^{-6} \left( \frac{M_{1}}{30M_{\odot}} \right)   M_{\odot}\rm yr^{-1}
\end{eqnarray}
\citep[see also][]{rpp05}. Each binary evolution calculation is 
terminated if the donor star has developed an iron core or the time steps exceed 30,000. 
We assume that the donor eventually collapses into a BH without any 
mass loss and kick, and the new born BH possesses negligible spin \citep[i.e. $ a_{\rm 2,BH} \sim 0$,][]{bkf20}. 
For simplicity, the spin of the accreting BH is assumed
to be aligned with the orbital angular momentum of the binary system ($ \theta_{1} = 0^{\circ}$). We follow \citet{p64} to treat  
the subsequent orbital evolution of the BBH system that is controlled by gravitational wave radiation.  
If the BBH system evolves to merge within a Hubble time, the effective spin parameter can be estimated to be 
\begin{eqnarray}
\chi_{\rm eff} \simeq \frac{M_{\rm 1,BH}}{M_{\rm 1,BH}+M_{\rm 2,BH}}a_{\rm 1,BH}.
\end{eqnarray}

\section{Results}

Figure 1 shows the evolution of a binary initially containing a $ 30M_{\odot} $ BH and a $ 80M_{\odot} $ zero-age main-sequence
companion in a 4-day orbit. The metallicity of the initial companion is taken to be 0.001.
The left and right panels correspond to the cases that the BH accretion rate is limited by 
$\dot{\mathcal{M}}_{\rm Edd} $ and $  \dot{M}_{\rm Edd} $, respectively. When the companion star
evolves to fill its Roche lobe, about $ 2M_{\odot} $ hydrogen envelope has been blown away due to a stellar wind. At this moment,
the donor star is still on main sequence (Case A evolution). A phase of rapid mass
transfer proceeds at a rate up to $ 10^{-2} M_{\odot}\rm yr^{-1}$, during which $ \sim 20M_{\odot} $
material is stripped from the donor star but the BH  hardly accretes. Subsequently, the system experiences a slow mass-transfer
phase that is driven by the nuclear-expansion of the donor star. During this phase, the mass transfer occurs at a rate of the order
$ 10^{-5} M_{\odot}\rm yr^{-1}$. In the $\dot{\mathcal{M}}_{\rm Edd} $ case, the BH can accrete $ \sim 8M_{\odot} $ material and reach
a spin of $ \sim 0.6$. 
After the Case A evolution, the system undergoes two other relatively short mass-transfer phases. 
Assuming that the donor directly collapses into a BH, we finally have 
a BBH system ($ M_{1, \rm BH}\sim 38M_{\odot} $ and $ M_{2, \rm BH}\sim 45M_{\odot} $) with an orbital period of $ \sim 1.8 $ days. 
About 0.6 Gyr later, this BBH is expected to merge with $ \chi_{\rm eff} \sim 0.3$.  It is obvious that 
the slow Case A mass-transfer phase is key to produce the high BH spin. This phase can last for
around 0.7Myr (see Figure 2). For comparison, in the $  \dot{M}_{\rm Edd} $ case,
the BH can totally accrete $ \sim 1M_{\odot} $ material with a relatively low spin of $ \sim 0.1$, and the
BBH merger has an effective spin of $ \chi_{\rm eff} \sim 0.06 $.

\begin{figure}[hbtp]
\centering
\includegraphics[width=0.5\textwidth]{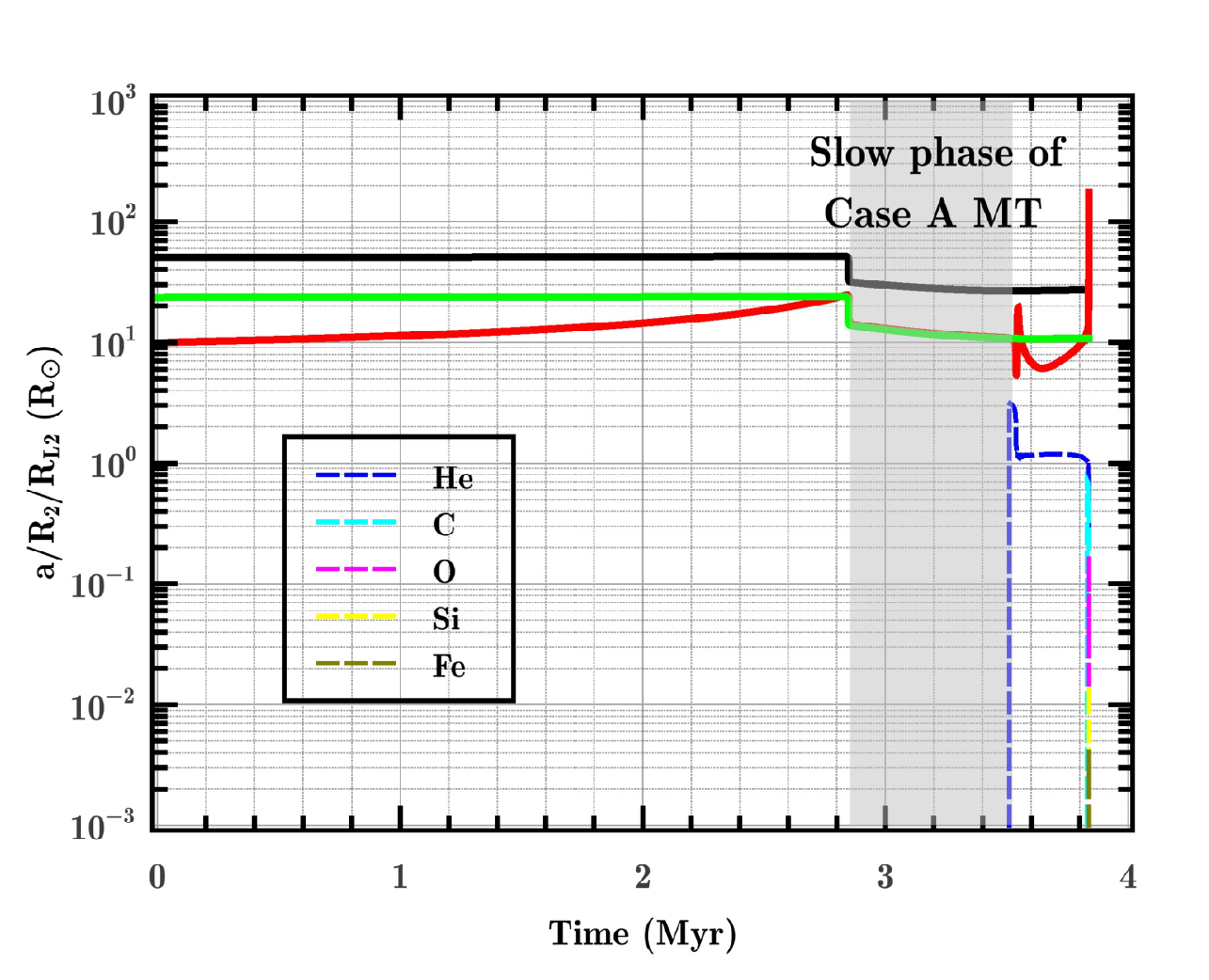}
\caption{Evolution of the donor radius $ R_{2} $ (the red curve), its Roche lobe radius $ R_{\rm L2} $ (the green curve) 
and the binary separation $ a $ (the black curve) as a function of time for a binary initially containing a $ 30M_{\odot} $ BH 
and a $ 80M_{\odot} $ donor in a 4-day orbit. The five dashed curves represent the radii of donor's cores.
The gray rectangle denotes the binary being in 
the slow phase of Case A mass transfer. At the time of $ \sim~3.83 $Myr, the stripped donor rapidly expands with radius 
beyond the size of the binary separation. At this moment, the donor has formed a $ \sim 43M_{\odot} $ core surrounded 
by a $ \sim 2M_{\odot} $ envelope at an advanced evolutionary stage. This system may have not enough 
time to develop a common-envelope phase before the donor collapses. Even if common-envelope evolution indeed occurs, this
binary can survive the spiral-in phase for $ \alpha_{\rm CE}\lambda \gtrsim 0.03 $ \citep[using Equations (1) and (2) in][]{sl14}.
   \label{figure1}}
\end{figure}

In Figure 3, we present the relations of $ M_{1, \rm BH} $ versus $ M_{2, \rm BH} $ (top panels) 
and $ M_{1, \rm BH} $ versus $ \chi_{\rm eff} $ 
(bottom panels) for the BBH systems that can merge within a Hubble time. The left and right panels correspond to 
the cases that the BH accretion rate is constrained by $\dot{\mathcal{M}}_{\rm Edd} $ and $  \dot{M}_{\rm Edd} $ (for comparison), respectively.
In each panel, the black and blue triangles correspond to the calculated results with $Z = 0.001 $ and $ Z=0.02 $, respectively.
Also plotted are the BBH mergers with high $ \chi_{\rm eff} $  measurements 
from gravitational wave observations \citep{aaa21b}. The initial binaries with a $20M_{\odot} $ donor always evolve
to produce wide BBHs that cannot merge within a Hubble time, so the triangle symbols do not appear to show these BBHs in this figure.
We can see that the calculated parameter distributions in the 
$\dot{\mathcal{M}}_{\rm Edd} $ case can roughly match the observations. In this case, the $ \chi_{\rm eff} $  distribution
is predicted to vary in a wide range of $ \sim 0 - 0.6 $. The source GW190517 seems to be an outlier with relatively light 
component masses ($ 25.3_{-7.3}^{+7.0} M_{\odot}$ and $ 37.4_{-7.6}^{+11.7} M_{\odot}$) and large effective spin 
($ \chi_{\rm eff} =0.52_{-0.19}^{+0.19} $). We note that mass transfer in the progenitor binaries with a $ 30M_{\odot} $
BH and a $ 40M_{\odot} $ companion
can eventually lead to the reversal of component masses (i.e. $ M_{\rm 1,BH} >  M_{\rm 2,BH} $). 
For one mass-reversed system, its calculated parameters
($ M_{2, \rm BH}\sim 18M_{\odot} $, $ M_{1, \rm BH}\sim 42M_{\odot} $ and $ \chi_{\rm eff} \sim 0.53 $) are consistent 
with the measured values of GW190517 within errors. 
In the $  \dot{M}_{\rm Edd} $ case, most of calculated BBH mergers have the
effective spins of $ \lesssim 0.2 $, which are obviously lower than observed.

\begin{figure*}[hbtp]
\centering
\includegraphics[width=0.85\textwidth]{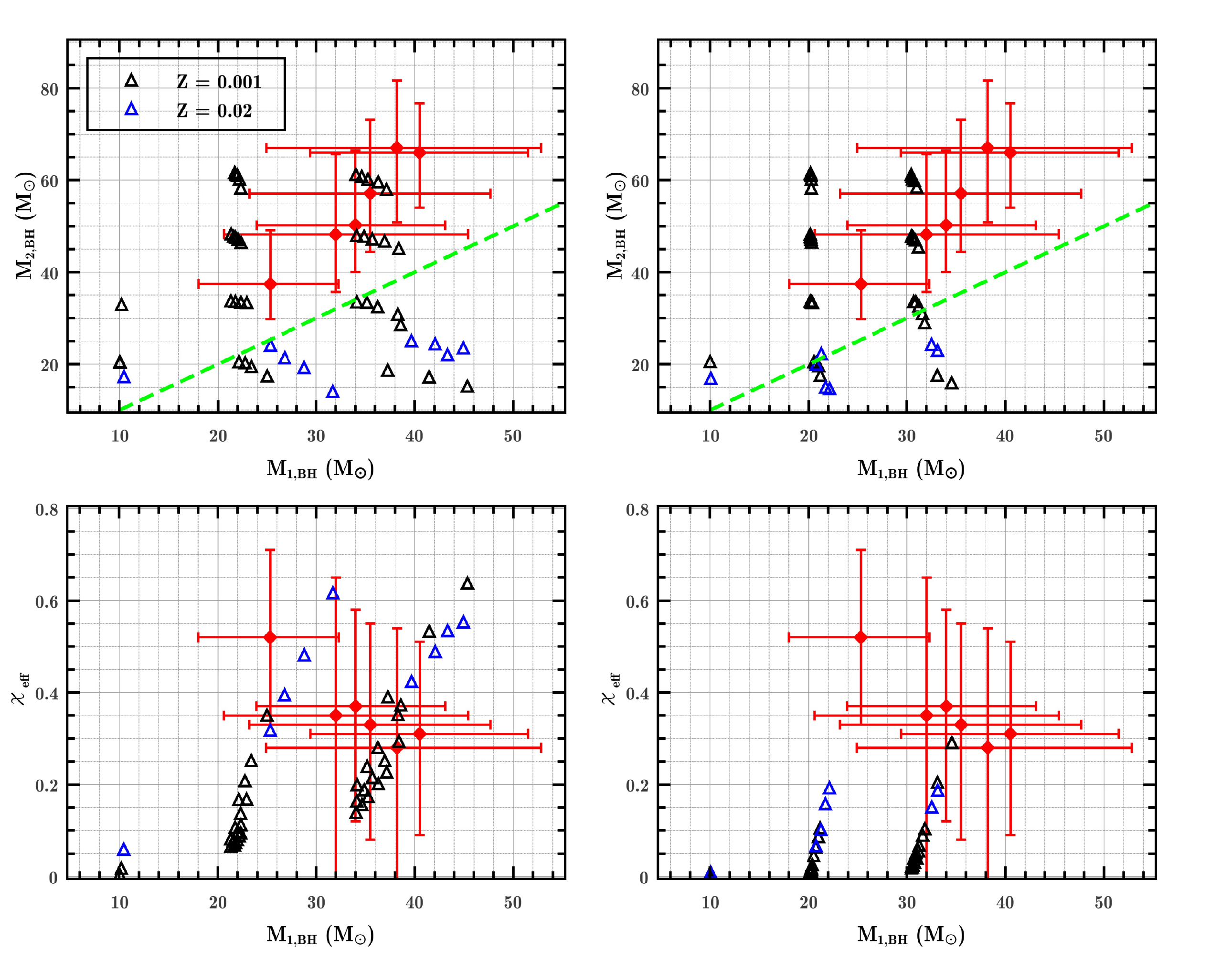}
\caption{Relations of $ M_{1, \rm BH} $ versus $ M_{2, \rm BH} $ (top panels) and $ M_{1, \rm BH} $ 
versus $ \chi_{\rm eff} $ (bottom panels) for calculated BBH mergers. 
The black and blue triangles correspond to the results with $Z = 0.001 $ and $ Z=0.02 $, respectively.
The red diamonds represent six observed systems with high $ \chi_{\rm eff} $ measurements,
corresponding to the sources GW170729, GW190517, GW190519, GW190620, GW190706, and GW190805 \citep{aaa21b}.  
The left and right panels correspond to the cases that the BH accretion rate is limited by 
$\dot{\mathcal{M}}_{\rm Edd} $ and $  \dot{M}_{\rm Edd} $ during the progenitor evolution, respectively. The green dashed line
shows the relation of $ M_{1, \rm BH}=M_{2, \rm BH} $. 
   \label{figure1}}
\end{figure*}

Figure 4 shows the relations between the effective spin parameters $ \chi_{\rm eff} $ of calculated BBH mergers and the initial orbital
periods  $P_{\rm i} $ of the progenitor binaries we evolved, assuming the BH accretion rate to be limited by $\dot{\mathcal{M}}_{\rm Edd} $. 
There is a tendency that the progenitor systems with shorter orbital periods
are more likely to produce the BBH mergers with higher effective spins. The binaries with $ P_{\rm i} \lesssim 5-10 $ 
days experienced relatively slow mass-transfer phases, leading to the formation of the BBH mergers with $  \chi_{\rm eff} \gtrsim 0.1-0.2 $.
Based on our calculations, all progenitor systems with $ M_{\rm 2i}/M_{\rm 1i} \leqslant 1 $ evolve to be wide BBHs 
that cannot merge within 
a Hubble time. On the other hand, the binaries with $ M_{\rm 2i}/M_{\rm 1i} > 3 $ and $P_{\rm i} \lesssim 5-10 $ days 
underwent dynamically unstable mass transfer \citep{sl21} and probably led to binary mergers involving a nondegenerate star and a BH. 
Hence the progenitor binaries with 
$ 1 \lesssim M_{\rm 2i}/M_{\rm 1i} \lesssim 3 $ and $ P_{\rm i} \lesssim 5-10 $ days mainly contribute the descendent BBH mergers with 
high effective spins. We note that some known BH binaries resemble the configurations of such progenitor systems. For example,
Cgy X-1 contains a BH of mass $ 21.2\pm 2.2M_{\odot} $ and a donor star of mass $ 40.6_{-7.1}^{+7.7}M_{\odot} $ in a 5.6-day 
orbit \citep{mbo21}. M33 X-7 hosts a $ 15.65\pm1.45 M_{\odot}$ BH and a $ 70.0\pm6.9 M_{\odot}$ donor orbiting around 
each other every 3.45 days \citep{omn07}. In the source LMC X-1, a  $ 10.91\pm1.41 M_{\odot}$ BH orbits a $ 31.79\pm3.48 M_{\odot}$
O-star companion every 3.9 days \citep{osm09}. These tight BH systems are expected to form from the primordial binaries that 
experienced either common envelope evolution \citep{omn07} or stable mass transfer \citep{vgf10,nvv21}. There is a caveat that the 
accreting BHs are assumed to initially have negligible spins in our calculations, which
seem to conflict with the observed high BH spins in these known binaries (\citealp{fk21}; see however \citealp{bdl21}). 

\begin{figure}[hbtp]
\centering
\includegraphics[width=0.5\textwidth]{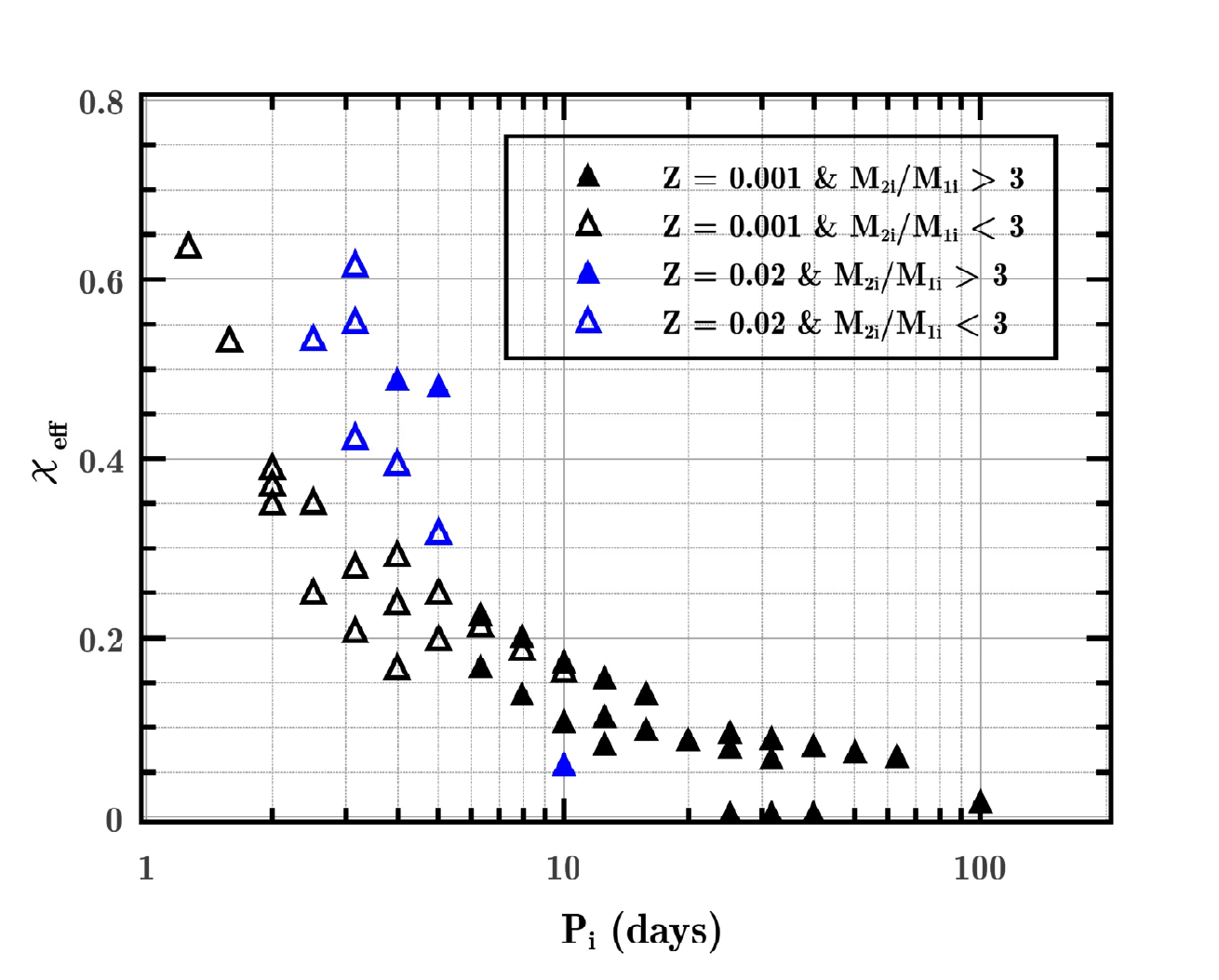}
\caption{Relations between the effective spin parameters of calculated BBH mergers and the initial orbital
periods of the progenitor binaries we evolved, by assuming that the BH accretion rate is limited by $\dot{\mathcal{M}}_{\rm Edd} $
during the evolution.
The open and filled triangles correspond to the progenitor systems with mass ratios (of the companion stars to the BHs) less than and larger
than 3, respectively.
   \label{figure1}}
\end{figure}

\section{Discussion}

We demonstrate how to produce BBH mergers with a fast-spinning component by way of Case A 
mass transfer. Formation of such mergers requires close progenitor binaries with a 
massive companion around a BH. They directly result from
close Wolf-Rayet star$ - $O star binaries \citep{vdh17}, who are widely observed in the Milky Way \citep{vdh01}
and some nearby galaxies \citep[e.g.,][]{sht16,ssh19}. These systems are the evolutionary products of primordial binaries
that experienced either a common envelope or a stable mass-transfer phase \citep[e.g.,][]{sl16,lss20}. A recent population
synthesis study including all binary evolutionary stages indicated that the stable mass-transfer channel dominates the formation
of local BBH mergers with large component masses \citep{vdc21}. Our work emphasizes that stable mass transfer from a massive
donor to a BH accretor is not only an important pathway for the formation of BBH mergers, but also likely responsible for the origin 
of the high spins of the binaries.

\acknowledgements
We thank the anonymous referee for constructive suggestions that helped 
improve this paper.
This work was supported by the Natural Science Foundation 
of China (Nos.~11973026, 12121003, 12041301 and U1838201), 
and the National Program on Key Research and 
Development Project (Grant No. 2021YFA0718500). 


\clearpage

\end{document}